\documentclass{aa}

\pdfoutput=1

\usepackage{graphicx}
\usepackage[varg]{txfonts}
\usepackage{amsmath}
\usepackage{verbatim}
\usepackage{latexsym}
\usepackage{multirow}
\usepackage{fixltx2e}
\usepackage{color}

\begin{document}

\title{Measuring the Wilson depression of sunspots using the divergence-free condition of the magnetic field vector}

\author{B. L\"optien\inst{1}
\and A. Lagg\inst{1}
\and M. van Noort\inst{1}
\and S.~K. Solanki\inst{1,2}}

\institute{Max-Planck-Institut f\"ur Sonnensystemforschung, Justus-von-Liebig-Weg 3, 37077 G\"ottingen, Germany
\and School of Space Research, Kyung Hee University, Yongin, Gyeonggi, 446-701, Republic of Korea}

\date{Received <date> /
Accepted <date>}

\abstract {The Wilson depression is the difference in geometric height of unit continuum optical depth between the sunspot umbra and the quiet Sun. Measuring the Wilson depression is important for understanding the geometry of sunspots. Current methods suffer from systematic effects or need to make assumptions on the geometry of the magnetic field. This leads to large systematic uncertainties of the derived Wilson depressions.}
{We aim at developing a robust method for deriving the Wilson depression that only requires the information about the magnetic field that is accessible from spectropolarimetry, and that does not rely on assumptions on the geometry of sunspots or on their magnetic field.}
{Our method is based on minimizing the divergence of the magnetic field vector derived from spectropolarimetric observations. We focus on large spatial scales only in order to reduce the number of free parameters.} 
{We test the performance of our method using synthetic Hinode data derived from two sunspot simulations. We find that the maximum and the umbral averaged Wilson depression for both spots determined with our method typically lies within 100~km of the true value obtained from the simulations. In addition, we apply the method to Hinode observations of a sunspot. The derived Wilson depression ($\sim 600$~km) is consistent with results typically obtained from the Wilson effect. We also find that the Wilson depression obtained from using horizontal force balance gives 110 \-- 180~km smaller Wilson depressions than both, what we find and what we deduce directly from the simulations. This suggests that the magnetic pressure and the magnetic curvature force contribute to the Wilson depression by a similar amount.}
{}

\titlerunning{Measuring the Wilson depression of sunspots}

\maketitle

\section{Introduction}
The geometric height at which unity continuum optical depth is reached is depressed within sunspots relative to the quiet Sun. This so-called Wilson depression~\citep{1774RSPT...64....1W} is caused by a lower opacity within the sunspot due to the lower temperature and a reduced gas pressure. The magnetic pressure and the curvature force of the strong magnetic field of sunspots balance this reduced gas pressure with the gas pressure of the surrounding quiet Sun. The connection between the Wilson depression and the strength and geometry of the magnetic field makes the Wilson depression an important quantity for understanding the structure of sunspots. In particular, it is not known by how much the curvature force of the magnetic field contributes to stabilizing the sunspot.

Unfortunately, the Wilson depression remains one of the more poorly known parameters of sunspots. Several studies have tried inferring the Wilson depression by making use of the horizontal force balance between the sunspot and the surrounding quiet Sun. However, since it is unknown by how much the curvature force contributes to the force balance, an accurate estimate of the Wilson depression with this method is not possible. Depending on the assumed influence of the curvature force, the derived Wilson depression lies in the range between 400-1000~km \citep{1993A&A...277..639S,1993A&A...270..494M,2004A&A...422..693M}.

The Wilson depression can also be estimated geometrically when the sunspot approaches the limb (i. e. via the Wilson effect). However, this method is influenced by radiative transfer or changes of the size of the umbra and penumbra with height \citep[see, e.g., the discussion in][]{2003A&ARv..11..153S}. Hence, the Wilson depression cannot be inferred very accurately when using this method. \citet{1972SoPh...26...52G} derived an average Wilson depression $z_{\rm W}$ of $600\pm 200$~km based on the Wilson effect. In contrast, \citet{1974SoPh...35..105P} measured a significantly larger Wilson depression of $950\--1250$~km with large spots having a higher Wilson depression ($z_{\rm W} = 1500\--2100$~km) than small spots ($z_{\rm W} = 700\--1000$~km). However, later results ($z_{\rm W} = 500 \-- 1000$~km) obtained by \citet{1983SoPh...88...71B} are in better agreement with those of \citet{1972SoPh...26...52G}.

Here we present an alternative method for measuring the Wilson depression that does not need to make any assumptions on the geometry of the magnetic field in sunspots or on the structure of the sunspot. Our method is based on imposing the divergence-free condition on the magnetic field vector deduced from the inversion of observed Stokes profiles. This approach has already been used by \cite{2010ApJ...720.1417P} to derive the small-scale corrugation of the $\tau=1$ layer of a small patch within the penumbra of a sunspot. Here, we modify this method to provide the large-scale corrugation of the $\tau=1$ layer within the entire sunspot. We first perform a test of the method: we use it to derive the Wilson depression from synthetic Hinode observations generated from two MHD simulations of sunspots \citep{2012ApJ...750...62R,2015ApJ...814..125R}. After that we apply it to one spot observed with Hinode. We also compare our results with the Wilson depression derived using the pressure method.

\section{Data}
\begin{figure*}
\centering
\begin{minipage}{5 cm}
\includegraphics[width=5 cm]{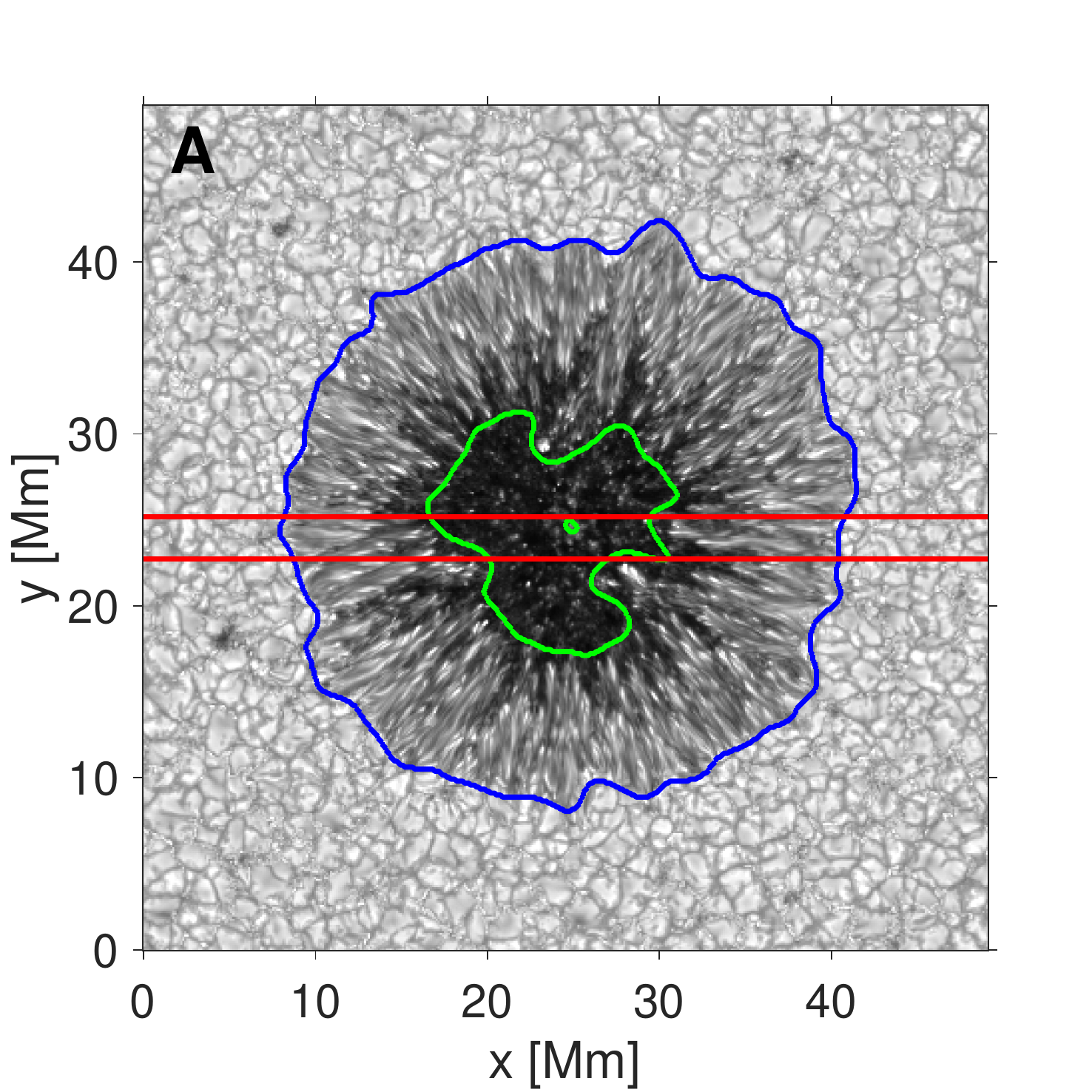}
\end{minipage}
\begin{minipage}{5 cm}
\includegraphics[width=5 cm]{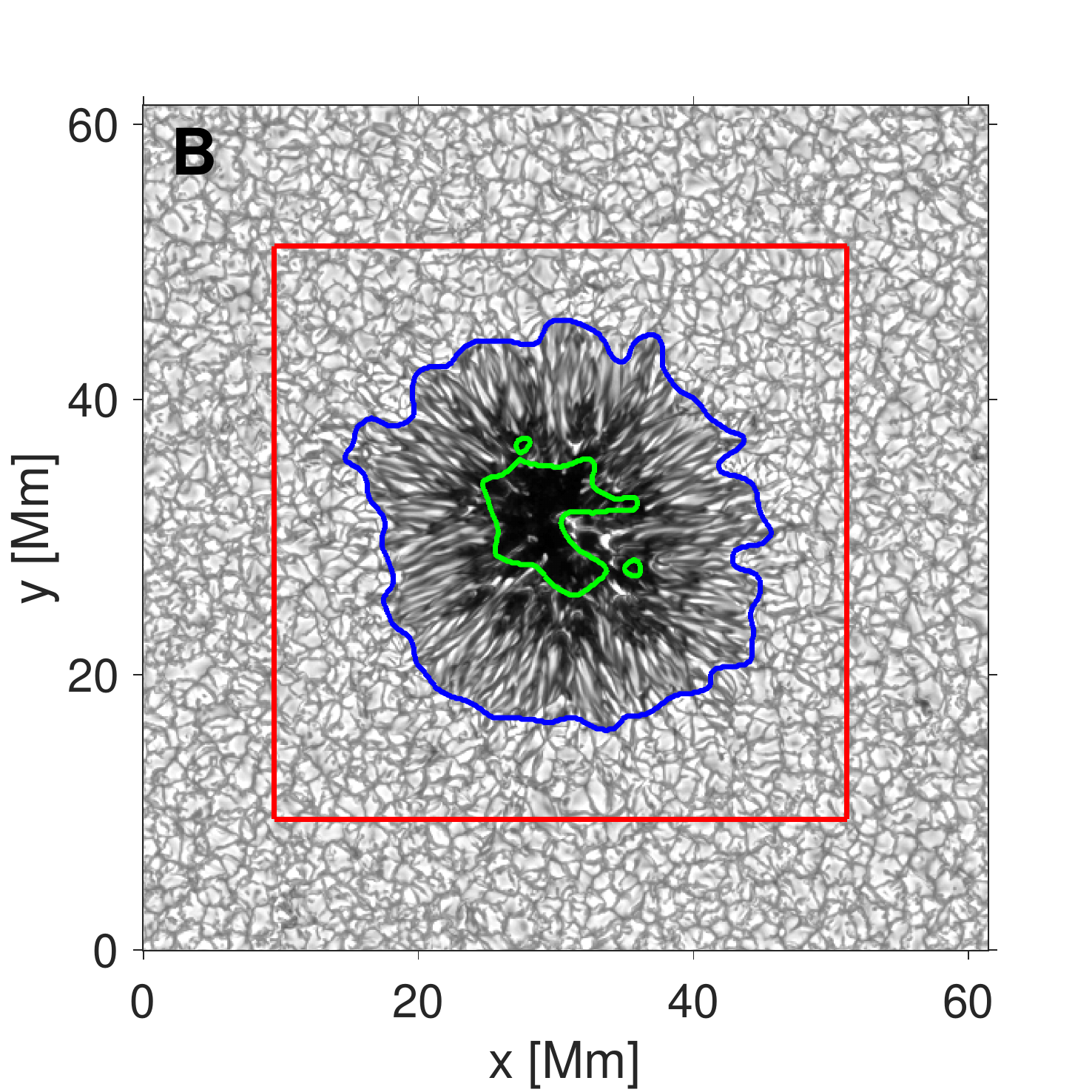}
\end{minipage}
\begin{minipage}{5 cm}
\includegraphics[width=5 cm]{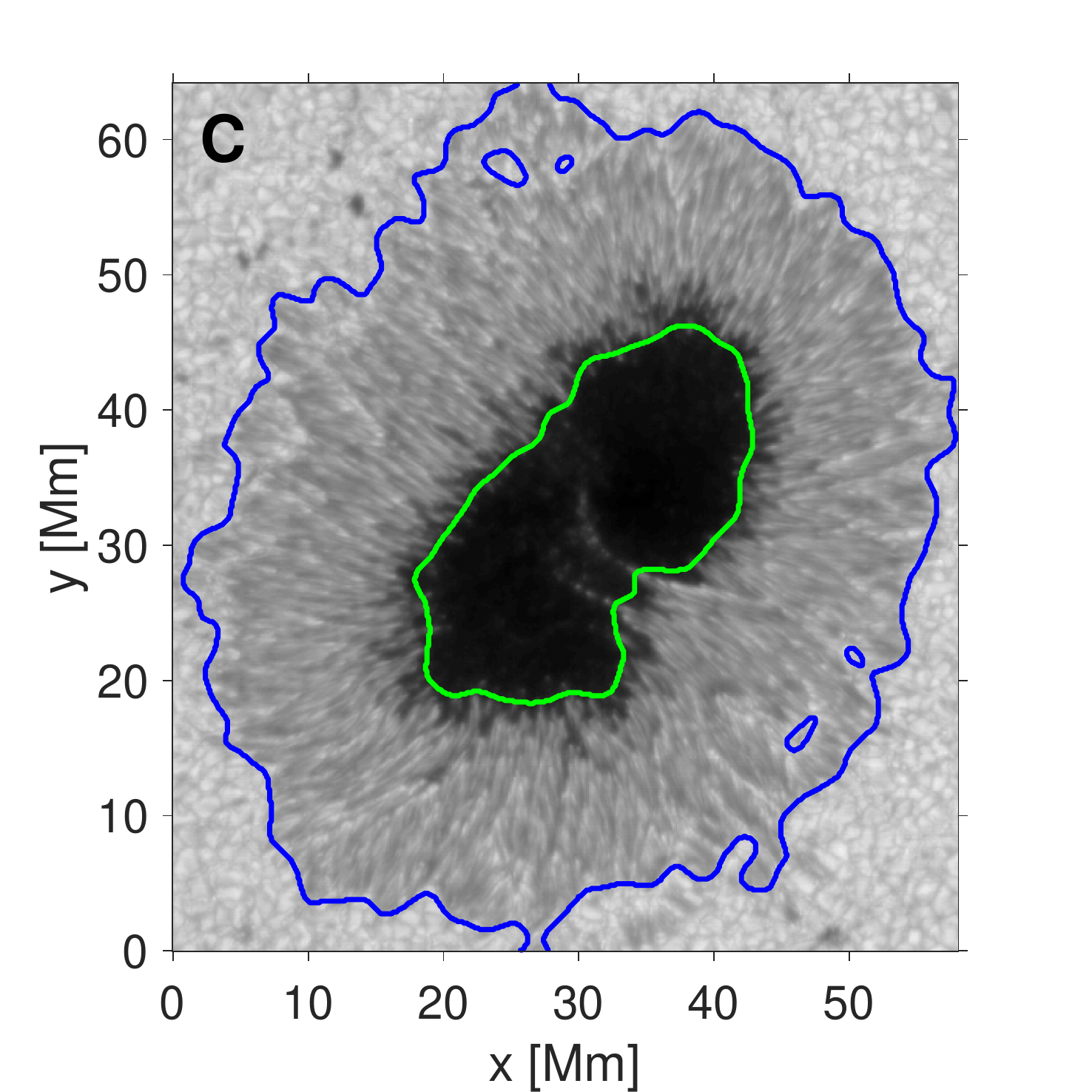}
\end{minipage}
\caption{Maps of the continuum intensity of the sunspots analyzed in this study. {\it Panels~A and~B} show sunspot simulations by \citet{2012ApJ...750...62R,2015ApJ...814..125R}, {\it panel~C} shows NOAA AR~10923 observed by Hinode SOT/SP on 14 November 2006. The {\it blue} and {\it green} contours indicate the outer and inner penumbral boundaries (defined as 30\% and 90\% of the continuum intensity level of the quiet Sun, after smoothing the continuum images with a 2D Gaussian with $\sigma = 812$~km). The {\it red boxes} outline the regions that we used for deriving the Wilson depression in the two MHD simulations. We compute the Wilson depression across the entire FOV for AR~10923.}
\label{fig:spots}
\end{figure*}

\begin{table*}
\caption{Parameters of the three spots that were analyzed in this study. The table contains the area of the spots, the maximum ($B_{\rm max}$) and the average ($B_{\rm av}$) magnetic field in the umbra (measured at $\log \tau = -0.9$), and the minimum ($T_{\rm min}$) and average ($T_{\rm av}$) temperature in the umbra (measured at $\log \tau = 0$).}
\label{tab:param}
\centering
\begin{tabular}{l l l l}
\hline\hline
Parameter & Simulation 1 & Simulation 2 & AR~10923\\
\hline 
area [Mm$^2$] & 938 & 650 & 2572 \\
$B_{\rm max}$ [Gauss] & 4247 & 5194 & 4341 \\
$B_{\rm av}$ [Gauss] & 3143 & 3624 & 2911\\
$T_{\rm min}$ [K] & 3896 & 3608 & 3579\\
$T_{\rm av}$ [K] & 4607 & 4325 & 4116\\
\hline
\end{tabular}
\end{table*}

\subsection{Synthetic Hinode data}
We generate synthetic Hinode data starting from two MHD simulations of sunspots by \citet{2012ApJ...750...62R,2015ApJ...814..125R} computed with the MURaM code~\citep{2005A&A...429..335V}. Simulation run 1 has a size of $49.152\times 49.152\times 2.048\ \textmd{Mm}^3$ with a resolution of $12\times 12\times 8\ \textmd{km}^3$ ($4096\times 4096\times 256$~pixels). Simulation run 2 has a size of $61.44 \times 61.44\times 2.976\ \textmd{Mm}^3$ with a resolution of $48\times 48\times 24\ \textmd{km}^3$ ($1280\times 1280\times 124$~pixels). Figure~\ref{fig:spots} shows continuum intensity images and Table~\ref{tab:param} lists some parameters of the spots (the area of the spots, the maximum and the average magnetic field in the umbra at $\log \tau = -0.9$, and the minimum and average temperature in the umbra at $\log \tau = 0$). We define the inner and outer boundary of the penumbra as 30\% and 90\% of the continuum intensity level of the quiet Sun, respectively, after smoothing the continuum images with a 2D Gaussian with $\sigma = 812$~km.

We compute line profiles of the pair of Fe I lines at $6301.5$~\AA \ and $6302.5$~\AA \ for all Stokes parameters from these simulations using the SPINOR code~\citep{2000A&A...358.1109F}; which uses the STOPRO routines for the forward calculation \citep{1987PhDT.......251S} for $\mu = 1$. In order to save computation time, we only compute the line profiles for a narrow strip across the sunspot for simulation run 1 (indicated by the red lines in the left panel in Figure~\ref{fig:spots}). For simulation run 2, we use the full field-of-view (FOV). We then degrade the data to the spectral resolution of Hinode SOT/SP and rebin the data to the pixel size of Hinode SOT/SP ($0.16''$). We do not convolve the data with the optical point-spread function (PSF) since an inversion with the spatially coupled version of SPINOR~\citep{2012A&A...548A...5V,2013A&A...557A..24V} would remove the influence of the PSF from the data anyway. Note that the sunspot AR~10923 observed by Hinode and analyzed here, was inverted with the spatially coupled version of SPINOR and we intend to apply this technique mainly to data inverted in the same manner. Finally, we add normally distributed noise to all four Stokes parameters with a signal-to-noise ratio that is proportional to the square root of the intensity and that reaches an average value of 1000 in the continuum.

\subsection{Hinode observations of AR 10923}
We use spectropolarimetric observations of AR~10923 made on 14 November 2006 in the Fe~I line pair at $6301.5$~\AA \ and $6302.5$~\AA \ from the spectropolarimeter on the Solar Optical Telescope \citep[SOT/SP,][]{2007SoPh..243....3K,2007ASPC..369..579L,2008SoPh..249..233I,2013SoPh..283..579L} onboard the Hinode spacecraft. At that time, this active region was located close to disk center ($x=66.8'',y=-114.4'',\mu =0.99$, see panel~C in Figure~\ref{fig:spots}). The SOT/SP provides the full set of Stokes parameters along both spectral lines with a pixel sampling of $\sim 0.16''$. This spot exhibits two filaments stretching far into the umbra, nearly splitting the umbra in two parts. This spot is significantly larger than the two simulated spots, however, the magnetic field in the umbra is weaker (see Table~\ref{tab:param}).

\section{Deriving the atmospheric conditions}
\begin{figure*}
\centering
\includegraphics[width=17cm]{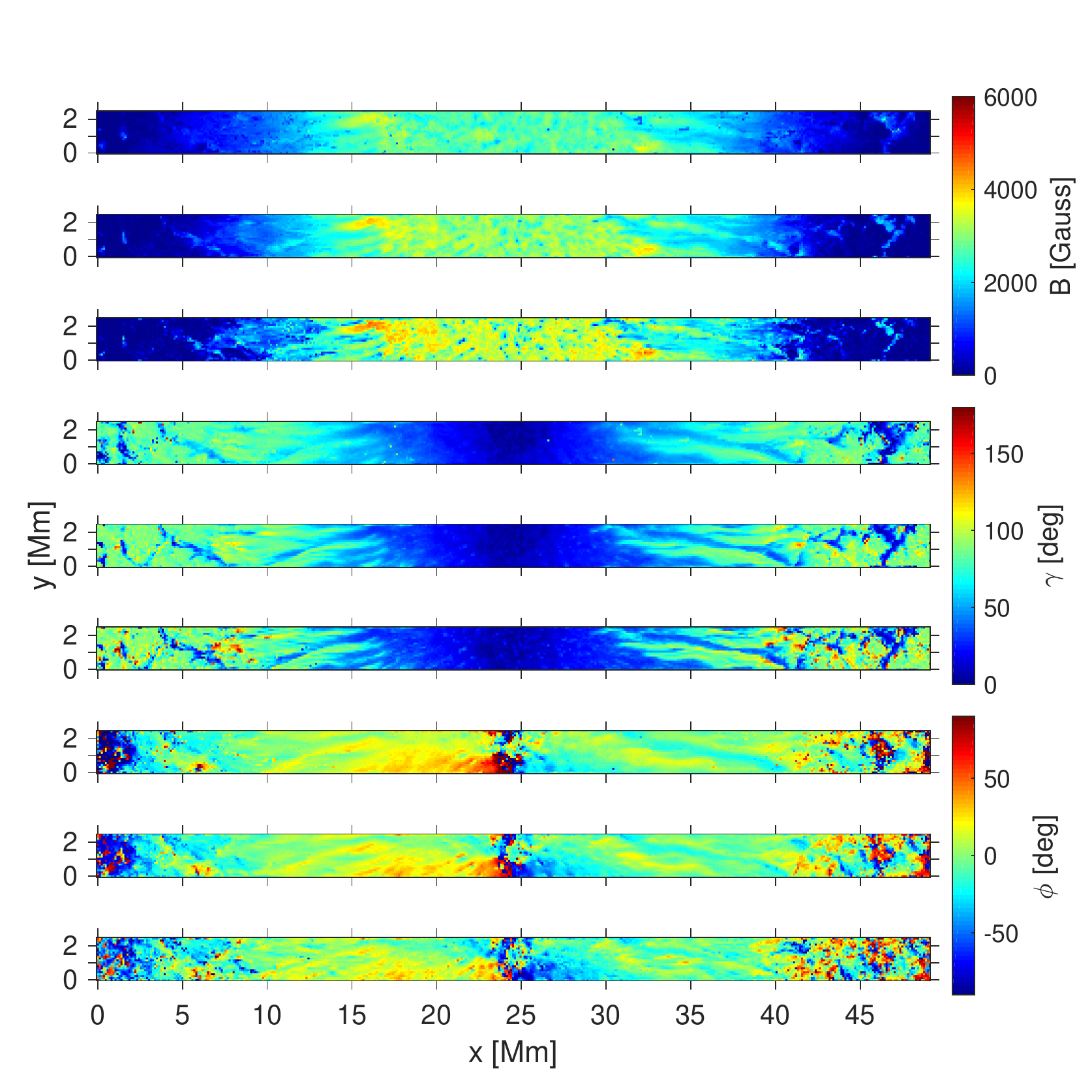}
\caption{Inverted magnetic field vector along the analyzed stripe through simulation run 1. The {\it top three rows} show the magnetic field intensity $B$, the {\it central three rows} show the inclination $\gamma$, and the {\it bottom three rows} the azimuth of the magnetic field $\phi$. We show each parameter at three different optical depths, {\it from top to bottom:} $\log \tau = -2.5,-0.9,0$.}
\label{fig:inv_spot1}
\end{figure*}

\begin{figure*}
\centering
\includegraphics[width=17cm]{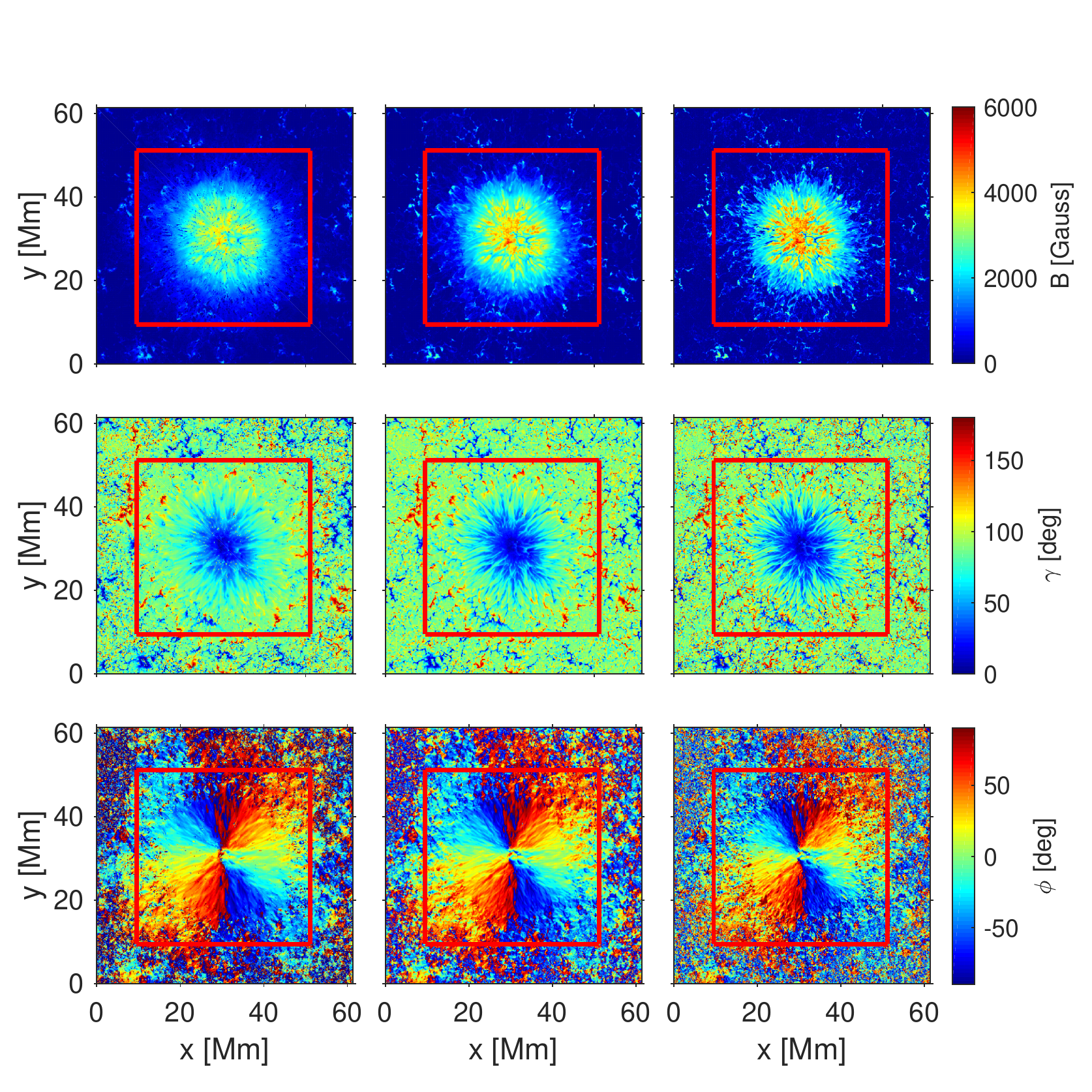}
\caption{Inverted magnetic field vector for simulation run 2. The {\it top row shows} the magnetic field intensity $B$, the {\it central row} show the inclination $\gamma$, and the {\it bottom row} the azimuth of the magnetic field $\phi$. {\it From left to right:} $\log \tau = -2.5,-0.9,0$. The {\it red box} highlights the region, where we derived the Wilson depression.}
\label{fig:inv_spot2}
\end{figure*}

\begin{figure*}
\centering
\includegraphics[width=17cm]{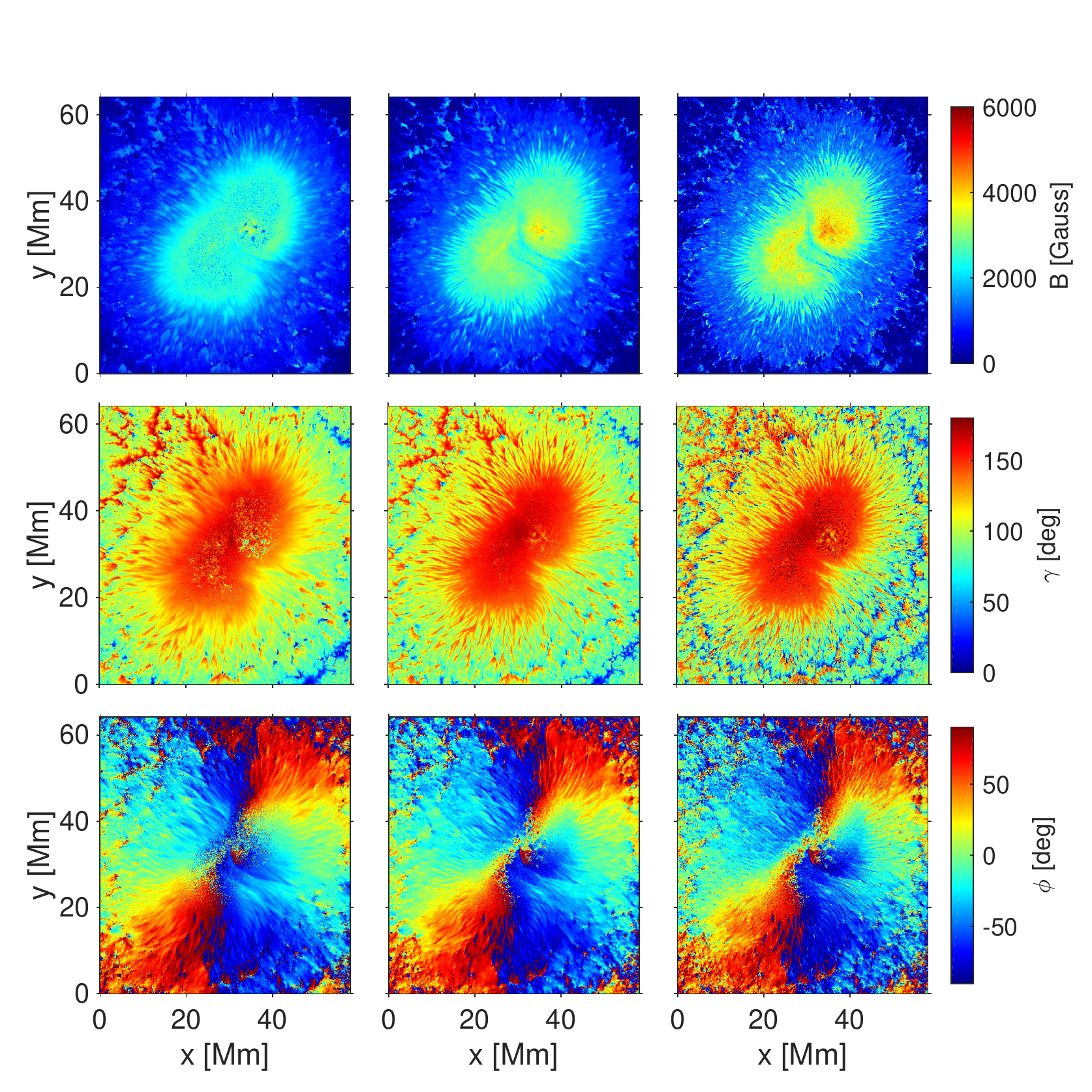}
\caption{Same as Figure~\ref{fig:inv_spot2} for the Hinode observation of AR~10923.}
\label{fig:inv_real_spot}
\end{figure*}

We derive the atmospheric parameters of both, the two simulated spots and AR~10923, by inverting the maps of the Stokes parameters with the SPINOR code under the assumption of local thermodynamic equilibrium (LTE). In case of the Hinode observations, we use the spatially coupled version of SPINOR \citep{2012A&A...548A...5V,2013A&A...557A..24V}. The synthetic data are inverted using the non-coupled version of SPINOR since we did not take into account the PSF when generating the synthetic data. For all three spots, we set three nodes in optical depth, placed at $\log{\tau} = -2.5,-0.9,0$. Figures~\ref{fig:inv_spot1} to \ref{fig:inv_real_spot} show the results of the inversion for the magnetic field strength $B$, the inclination $\gamma$, and the azimuth $\varphi$ for all the spots at these optical depths. Using a spline interpolation, we then remap the results of the inversion on an equidistant grid in $\log{\tau}$ ranging from $-6$ to $+1.5$ with a sampling of $0.1$.

Afterward, we resolve the $180^\circ$ azimuthal ambiguity by using the Non-Potential Magnetic Field Computation method \citep[NPFC,][]{2005ApJ...629L..69G}. We can test the performance of the NPFC code with the help of the synthetic data. In most cases, the code resolves the ambiguity correctly within the sunspot. However, the code sometimes fails in regions where the magnetic field vector is almost exactly parallel to the line-of-sight. This occurs predominantly at optical depths where the inversion is less accurate. We address this issue by applying the NPFC code only to the data at $\log{\tau} = -0.9$. We then successively resolve the ambiguity at greater and lower heights by demanding the magnetic field vector to vary smoothly with height. We select the solution of the ambiguity, where the magnetic field vector is the closest to the one in the adjacent layer, where we have already resolved the ambiguity. This leads to an accurate disambiguation within the sunspot. In the surroundings of the sunspots, however, the NPFC code fails in many cases to find the correct solution, even at $\log{\tau} =-0.9$.

\section{Measuring the Wilson depression}
\subsection{The method}\label{sect:method}
We determine the geometric height of the $\tau = 1$ layer of the sunspot in a similar way as introduced by \cite{2010ApJ...720.1417P}, who could infer the geometrical height of the $\tau = 1$ surface in an inverted atmosphere of a small patch of the penumbra of a sunspot observed with Hinode. Their method is based on the divergence-free condition of the magnetic field vector and on ensuring force balance. These two conditions are not fulfilled in an inverted atmosphere since the unknown height of the $\tau = 1$ layer causes an offset in the geometric height scale of the individual pixels. However, both the divergence of the magnetic field and the deviations from force balance are minimized when shifting the atmosphere at each pixel by the corresponding height of the $\tau = 1$ surface. Hence, this approach allows to infer the Wilson depression. \cite{2010ApJ...720.1417P} use the following merit function:
\begin{align}
\chi^2 = \sum_{m,n} w_1 |\vec{F}|^2 +w_2 (\vec{\nabla} \cdot \vec{B})^2
\end{align}
The sum is over all the pixels in the images (indicated by the indices $m$ and $n$). The first term consists of the residual force $\vec{F} = \vec{J} \times \vec{B} + \rho \vec{g} - \vec{\nabla}P_g$. Here, $\vec{J}$ is the current density, $\vec{B}$ is the magnetic field vector, $\rho$ is the density, $\vec{g}$ is the surface gravity of the Sun, and $P_g$ is the gas pressure. The influence of flows on the force balance is neglected. The second term in the merit function is the divergence of the magnetic field. The relative contributions of the two terms are balanced by the coefficients $w_1$ and $w_2$.

The magnetic field vector and all other atmospheric parameters in the merit function depend on the geometric height $z$. This height scale has two contributions. The first one is the $z$-scale $z_{\rm rel}$ of the stratification relative to the $\tau = 1$ layer of each pixel, which is provided by the inversion. The second one is the height of the $\tau = 1$ layer at each pixel (i. e. the Wilson depression $z_{\rm W}$), which we want to determine:
\begin{align}
z(x,y,\tau)=z_{\rm rel}(x,y,\tau) + z_{\rm W}(x,y)
\end{align}
Changing $z_{\rm W}$ at the individual pixels corresponds to shifting the inverted atmosphere vertically. The merit function has its minimum when the inverted atmospheres for all pixels are aligned with respect to each other, i. e., when $z_{\rm W}(x,y)$ is the corrugation of the $\tau = 1$ surface. The alignment of the individual atmospheres can be evaluated at different heights $z_{\rm rel}$ relative to the $\tau = 1$ surface. \cite{2010ApJ...720.1417P} used $z_{\rm rel} = 200$~km and set $z_{\rm W} = 0$ as the height used to compute the merit function for an arbitrary pixel. They used a genetic algorithm for minimizing the merit function.

Here, we want to extend the method of \cite{2010ApJ...720.1417P} to derive the corrugations of the $\tau = 1$ layer of the entire spot. This is not straightforward since the Wilson depression at each pixel is a free parameter in the merit function. So, the number of free parameters becomes extremely large when applying this method to the entire spot. We reduce the number of free parameters by restricting the analysis to the large-scale corrugation of the $\tau = 1$ layer. An efficient way to achieve this is to write the Wilson depression $z_{\rm W}(x,y)$ as a Fourier series of the horizontal wavenumbers $k_x$ and $k_y$ and then drop the terms above a maximum wavenumber in the Fourier series of $z_{\rm W}$, therefore removing the small-scale corrugations. Now, the number of free parameters does not correspond to the number of pixels anymore, but it depends on the maximum wavenumber that is considered. This approach allows us to reduce the number of free parameters in the merit function significantly without affecting the determination of the large-scale Wilson depression.

However, neglecting small-scale corrugations of the $\tau = 1$ layer affects the divergence of the magnetic field, especially on small spatial scales. We address this problem by writing the merit function in Fourier space using Parseval's theorem:
\begin{align}
\chi^2 = \sum_{m,n} (\vec{\nabla} \cdot \vec{B})^2 = \frac{1}{N_x N_y} \sum_{k,l} (\mathcal{F} (\vec{\nabla} \cdot \vec{B}))^2
\end{align}
Here, the symbol $\mathcal{F}$ indicates a 2D discrete Fourier transform (in the $x$- and in the $y$-direction), defined as
\begin{align}
\hat{a}_{k,l} = \sum_{m=0}^{N_x-1}\sum_{n=0}^{N_y-1} e^{-2\pi i \left ( \frac{mk}{N_x} + \frac{nl}{N_y} \right )} \cdot a_{m,n}.
\end{align}
The parameters $N_x$ and $N_y$ are the number of pixels along the $x$-axis and the $y$-axis, respectively. The indices $k$ and $l$ indicate the dimensionless wavenumber indices and are connected to the physical wavenumber by the relation $k_x = \frac{2\pi}{N_x \Delta x}\cdot k$ in case of the $x$-direction with $\Delta x$ being the spatial resolution. Contrary to \cite{2010ApJ...720.1417P}, we do not include the force terms in our merit function. In our case, including the force terms leads to a less accurate determination of the Wilson depression. This is caused by neglecting advection ($\rho \left ( \vec{v} \cdot \vec{\nabla} \right ) \vec{v}$) in the force balance (spectropolarimetry can only measure the line-of-sight velocity, not the full velocity vector). In the penumbra of the two simulated spots, the advection term reaches a similar magnitude as the Lorentz force due to the Evershed and the reverse Evershed flow. Hence, ignoring advection causes the force term in the merit function not to be sensitive to the large-scale Wilson depression anymore.

All terms in the merit function are positive, meaning that the divergence has to be zero at all spatial scales in order to minimize the merit function. So, one can restrict the analysis to the large spatial scales of the divergence only, which are less affected by neglecting the small-scale corrugations of the $\tau = 1$ layer. Both for the Wilson depression and for the divergence of the magnetic field vector, we only consider spatial scales up to maximum dimensionless wavenumbers $k_{\rm max}$ and $l_{\rm max}$. This decreases the effective spatial resolution significantly (given by the wavelength corresponding to the indices $k_{\rm max}$ and $l_{\rm max}$). When using a maximum dimensionless wavenumber $k_{\rm max} = l_{\rm max} = 3$, for example, we retrieve an effective spatial resolution of 14 Mm for the part of simulation run 2 that is highlighted by the red box in Figure 1.

We filter both, the Wilson depression and the magnetic field vector in Fourier space to include only the information about large spatial scales. Hence, a map of the large-scale magnetic field at a fixed geometric height is affected by the filtering in two ways. Ignoring the small-scale corrugations of the $\tau = 1$ layer causes an error in the derived magnetic field vector if the magnetic field varies strongly with height at a given pixel. Afterwards, variations of the magnetic field itself on small spatial scales are removed. This filtering should remove most of the errors introduced by neglecting the small-scale corrugations of the $\tau = 1$ layer. 

This assumption can be tested using the simulated sunspots. We compute maps of all three components of the magnetic field vector at a fixed geometrical height (200~km above the mean height of the $\tau = 1$ layer) from simulation run 2 for two different cases. In the first case, we derive the magnetic field on a grid in geometric height that includes small-scale corrugations of the $\tau = 1$ surface, in the second case, we neglect corrugations above a maximum dimensionless wavenumber $k_{\rm max} = l_{\rm max} = 3$. Afterwards, we also remove the signal at spatial scales above $k_{\rm max}$ or $l_{\rm max}$ from the maps of the magnetic field vector. The correlation coefficient between these two sets of maps of the magnetic field vector is $0.98$ for all three components of the magnetic field vector, indicating that the small-scale corrugations of the $\tau = 1$ surface do not have a strong influence on the large-scale magnetic field. Using this approach leads to the following expression for the merit function:
\begin{align}
\chi^2 &= \frac{1}{N_x N_y} \sum_k^{k_{\rm max}} \sum_l^{l_{\rm max}} (\mathcal{F} (\vec{\nabla} \cdot \vec{B}))^2\\
&= \frac{1}{N_x N_y} \sum_k^{k_{\rm max}} \sum_l^{l_{\rm max}} \left ( i k_x \hat{B}_x + i k_y \hat{B}_y + \mathcal{F} \left ( \frac{\partial B_z} {\partial z} \right ) \right ) ^2 \label{eq:1}
\end{align}
Here, $\hat{B}_x(k_x,k_y,z)$ and $\hat{B}_y(k_x,k_y,z)$ are the Fourier transforms of $B_x(x,y,z)$ and $B_y(x,y,z)$, respectively. Again, the geometric height $z$ consists of the relative height above the $\tau = 1$ surface $z_{\rm rel}$ and the Wilson depression $z_{\rm W}$. We evaluate the merit function around a fixed reference height $z_{\rm rel}$ and require the derived Wilson depression to have a mean value of zero. Like \cite{2010ApJ...720.1417P}, we minimize our merit function using a genetic algorithm. We run the genetic algorithm ten times and use the solution which minimizes the merit function the most. After running the genetic algorithm, we define $z_{\rm W}=0$ as the height of the $\tau = 1$ surface averaged over the part of the FOV that has a distance of at least 7~Mm to the sunspot. This minimum distance is necessary because the height of the $\tau = 1$ surface can be depressed in the close surroundings of the sunspot.

Our method relates the magnetic field vector in the quiet Sun and in the umbra at the same geometric height. This requires an accurate measurement of the magnetic field vector over a broad range in height since sunspots have a Wilson depression of a few 100~km. So, in order to derive reliable estimates of the Wilson depression, the magnetic field vector retrieved by the inversion needs to be reliable over a broad range in height around the reference height $z_{\rm rel}$. Errors in the inversion occur predominantly at high optical depths (we do not consider data at heights $z_{\rm rel} < 85$~km) and at very low optical depths. We choose $z_{\rm rel}$ to be as small as possible while ensuring that all inferred heights are greater than our threshold of 85~km. The inversion also suffers from systematic errors, such as the assumption of hydrostatic equilibrium. In the simulation data, this is a good approximation in granules and in the umbra but not so good in the intergranular lanes and in the penumbra. Errors occurring on small spatial scales should have limited influence though, due to our focus on large spatial scales.

We perform our analysis in Fourier space. This means that the magnetic field vector and the derived Wilson depression have periodic boundary conditions when expressing them in Fourier space. This is a good approximation when the field of view also contains the surroundings of the sunspot, where the magnetic field and the Wilson depression are significantly lower than within the spot.

\begin{table*}
\caption{Wilson depression of the three spots that were analyzed in this study. The first column indicates the spot, the second column the method that was used to infer the Wilson depression. The following columns are explained in the text in detail.}
\label{tab:z_W}
\centering
\begin{tabular}{l l l l l l l l}
\hline\hline
Spot & Method & $z_{W,max}$ [km] & $z_{W,U}$ [km] & $\Delta z_{W,U}$ [km] & $z_{W,PU}$ [km] & $\Delta z_{W,PU}$ [km] & $\Delta z_{W,QS}$ [km]\\
\hline 
Simulation 1 & True solution (full res.) & 719 & 552 & & 286 & & \\
	     & True solution (filtered)  & 606 & 546 & & 284 & & \\
	     & Divergence                & 702 & 539 & 58 & 264 & 57 & 22 \\
	     & pressure (full res.) & 598 & 424 & 128 & 163 & 124 & 33\\
	     & pressure (filtered)  & 482 & 418 & 129 & 167 & 118 & 23\\
\hline
Simulation 2 & True solution (full res.) & 880 & 602 & & 253 & \\
	     & True solution (filtered)  & 644 & 559 & & 252 & \\
	     & Divergence                & 591 & 478 & 95 & 296 & 81 & 47\\
	     & pressure (full res.) & 791 & 489 & 115 & 128 & 126 & 34\\
	     & pressure (filtered)  & 548 & 449 & 111 & 130 & 123 & 22\\
\hline
AR 10923     & Divergence       & 720 & 600 & & 195 & & \\
	     & pressure (full res.) & 742 & 427 & & 82 & & \\
	     & pressure (filtered)  & 555 & 414 & & 82 & & \\
\hline
\end{tabular}
\end{table*}

\subsection{Synthetic Hinode data}\label{sect:simulated_spots}
\begin{figure*}
\centering
\includegraphics[width=17cm]{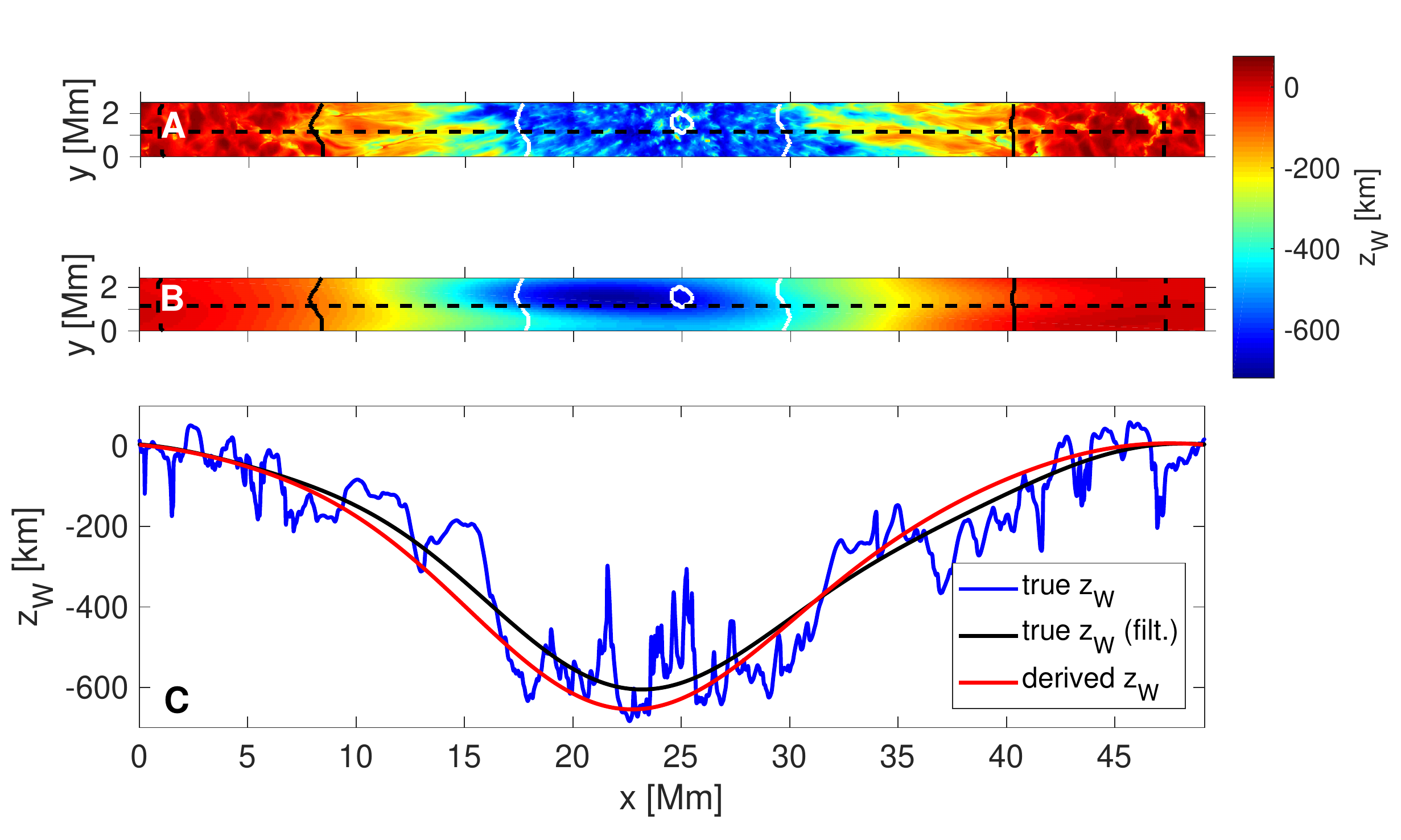}
\caption{Wilson depression of simulation run 1. {\it Panel~A} shows the true Wilson depression, {\it panel~B} the Wilson depression derived from the inversion. The {\it black} and {\it white} contours indicate the outer and the inner penumbral boundary. The {\it dashed black contour} shows the region that is used for defining $z_{\rm W} = 0$ (distance of at least 7~Mm to the sunspot). {\it Panel~C} shows the Wilson depression along a cut across the sunspot (marked by the {\it dashed lines}in {\it panels~A and B}). {\it Red:} Wilson depression derived from synthetic observations, {\it blue:} true Wilson depression, {\it black:} true Wilson depression filtered in Fourier space to the same range in wavenumber as the derived solution from the inversion, for direct comparison with the {\it red curve}.}
\label{fig:WD_spot1}
\end{figure*}

\begin{figure*}
\centering
\includegraphics[width=17cm]{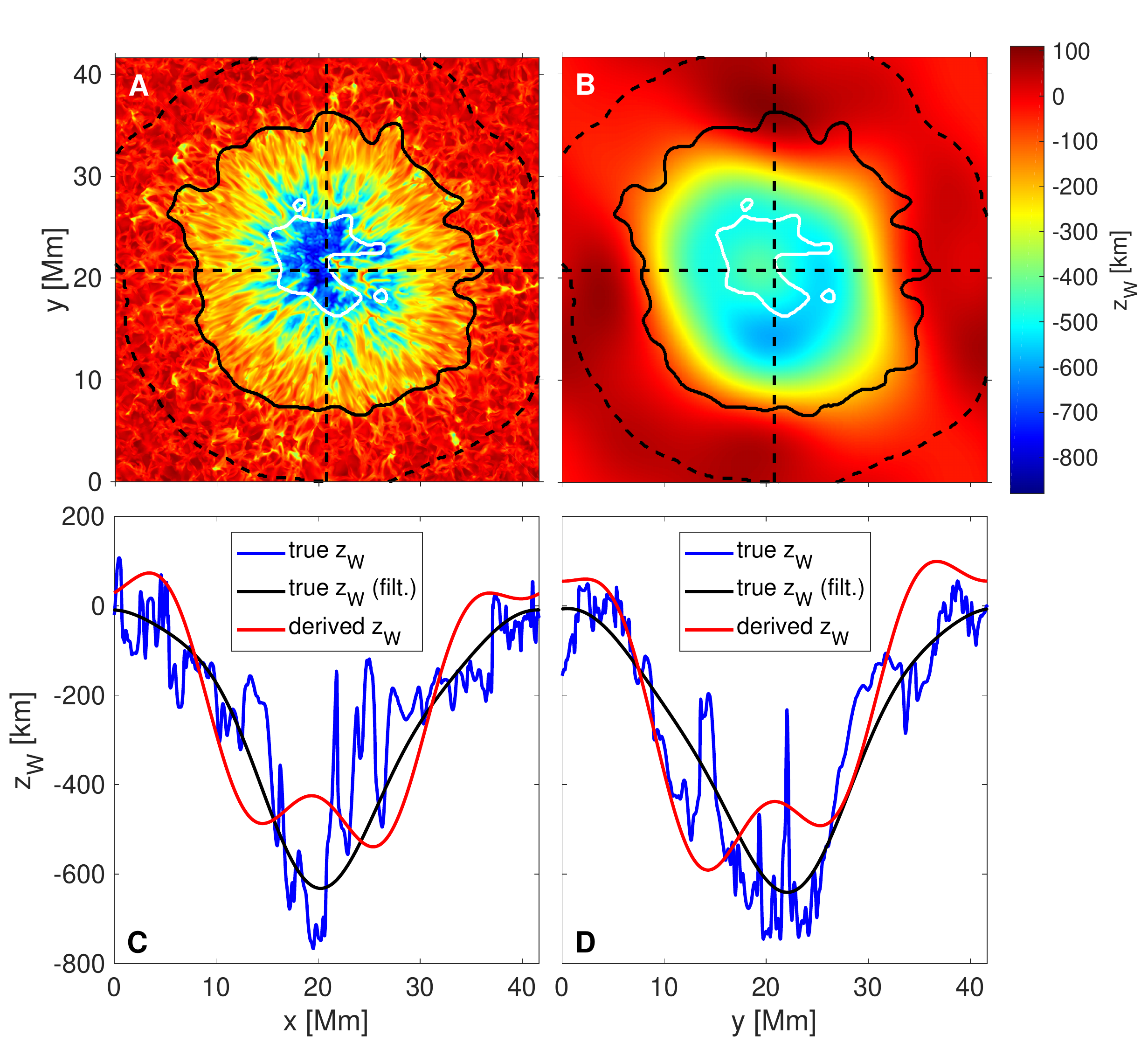}
\caption{Wilson depression of simulation run 2. {\it Panel~A} shows the true Wilson depression, {\it panel~B} the Wilson depression derived from the inversion. The {\it black} and {\it white} contours indicate the outer and the inner penumbral boundary. The {\it dashed black contour} shows the region that is used for defining $z_{\rm W} = 0$ (distance of at least 7~Mm to the sunspot). The {\it bottom row} shows the Wilson depression along cuts across the sunspot (marked by the {\it dashed lines} in the {\it top row}). {\it Panel~C} shows a cut along the $x$-direction, {\it panel~D} shows a cut along the $y$-direction. {\it Blue:} true Wilson depression, {\it black:} true Wilson depression filtered in Fourier space to the same range in wavenumber as the derived solution from the inversion (shown in {\it red}).}
\label{fig:WD_spot2}
\end{figure*}

Figures~\ref{fig:WD_spot1} and~\ref{fig:WD_spot2} show the Wilson depressions derived from synthetic Hinode data for the two simulated sunspots.  For comparison, we also show the true values of the Wilson depression for both spots which were inferred by resampling the MHD cubes on a grid in optical depth using SPINOR. We set $z_{\rm W} = 0$ as the height of the $\tau = 1$ surface averaged over the region of the FOV that has a distance of at least 7~Mm to the outer boundary of the penumbra (see dashed black contours in Figures~\ref{fig:WD_spot1} and~\ref{fig:WD_spot2}). In Table~\ref{tab:z_W}, we quantitatively compare the Wilson depression inferred by our method with the true solution and results from the pressure method (see Sect.~\ref{sect:force}). We list the maximum Wilson depression ($z_{\rm W,max}$) and its average over the umbra ($z_{\rm W,U}$) and over the penumbra ($z_{\rm W,PU}$), both for the full spatial resolution (called ``full res.'' in the table) and after degrading it to the resolution of our method (``filtered'' in the table). We also derive error estimates of the Wilson depression for the umbra ($\Delta z_{\rm W,U}$), the penumbra ($z_{\rm W,PU}$), and the area outside the sunspot ($\Delta z_{\rm W,QS}$). We derive these errors from maps of the absolute value of the difference between our solutions and the true Wilson depression (after degrading it to the spatial resolution of our derived Wilson depressions). We then define the errors to be the average of this difference map over the respective region of the spot.

For spot 1, we compute the Wilson depression only along a slice across the sunspot to limit the computation time due to the large number of pixels. Within this slice, the magnetic field vector is not periodic in the $y$-direction, in particular the $B_y$ component. Since this non-periodic boundary condition would affect the filtering in Fourier space, we compute the divergence using the full spatial resolution in the $y$-direction. However, we do apply Fourier filtering in the $x$-direction. We express the Wilson depression in Fourier space, though, in order to reduce the number of free parameters ($k_{\rm max} = 3$, $l_{\rm max} = 1$). The main differences between our solution and the true Wilson depression occur on small spatial scales, which we do not consider in our method. We degrade the true solution to the same spatial resolution as our derived Wilson depression (black curve in the bottom panel of Figure~\ref{fig:WD_spot1}) in order to remove the influence of the differences in spatial resolution. Our derived Wilson depression is generally in good agreement with the degraded true solution, but the maximum Wilson depression is $\sim 100$~km larger than the degraded true solution (see Table~\ref{tab:z_W}). The averages of the Wilson depression over the umbra and the penumbra are in good agreement, though. In both cases, the mean absolute difference between our solution and the true Wilson depression is of the order of 60~km.

This error estimate is affected both by uncertainties introduced by the inversion of the Stokes parameters and by the stability of the genetic algorithm. We evaluate the stability of the minimization by computing the standard deviation of the Wilson depression inferred by running the genetic algorithm ten times. The resulting errors (34~km for the maximum Wilson depression and 17~km for the average over the umbra) are significantly smaller than the difference between the derived and the true Wilson depression. This suggests that the error is dominated by uncertainties in the inversion.

Spot 2 exhibits a very complex Wilson depression in the umbra with changes of several 100~km on small spatial scales (see Figure~\ref{fig:WD_spot2}). These small-scale corrugations cannot be resolved by our method (here we use $k_{\rm max} = l_{\rm max} = 3$). Again, for better comparison, we smooth the true solution in order to account for this effect. Afterwards, both the maximum Wilson depression and its averages over the umbra and the penumbra are in good agreement (see Table~\ref{tab:z_W}). A remarkable difference between our solution and the true solution occurs at $x\approx 24 \ \textmd{Mm}, \ y\approx 20$~Mm. At this position, there is a filament stretching to the umbra and the $\tau = 1$ layer is located a few 100~km higher than in the surrounding umbra. Our method finds an extremely large Wilson depression at this location, in contradiction to the true solution. This error in the derived Wilson depression is probably caused by uncertainties in the inversion. Along the filament, the magnetic field changes strongly with height (by up to 2000~Gauss over a range of 100~km at around $\tau = 1$). This height dependency cannot be described correctly by the inversion. This feature not only affects the Wilson depression locally but also in the surrounding umbra. This is because the spatial resolution of our method is very low and because the Fourier transform is non-local (a perturbation of the signal at a given point also affects distant points after filtering in Fourier space). Again, we compute the mean absolute value of the difference between our derived Wilson depression and the true solution. The differences are roughly a factor $1.5$ larger than the ones for spot 1 and much larger than the uncertainty introduced by the genetic algorithm (13~km for the maximum Wilson depression and 9~km for the average over the umbra), in agreement with the findings from spot 1.

The inferred Wilson depression depends on the parameters of the merit function. Apart from the Wilson depression, the merit function (Eq. \ref{eq:1}) depends on the range of spatial scales that is considered (given by $k_{\rm max}$ and $l_{\rm max}$) and the height above $\tau = 1$ ($z_{\rm rel}$) where the merit function is evaluated. As stated in the previous section, the height $z_{\rm rel}$ is a crucial parameter since it determines, what range in height (and optical depth) is used for inferring the Wilson depression. We choose the reference height $z_{\rm rel}$ to be as small as possible while ensuring that all inferred heights are greater than our threshold of 85~km (350~km for spot 1, 300~km for spot 2). For larger reference heights, the derived Wilson depression becomes less reliable. We also restrained the solution to low wavenumbers. Larger wavenumbers should be avoided since in that case the solution for the Wilson depression is more likely to exhibit an oscillatory behavior, especially in the quiet Sun.

\subsection{Comparison with the pressure method}\label{sect:force}
For comparison, we also derive the Wilson depression by demanding horizontal force balance between the sunspot and the surrounding quiet Sun \citep{1993A&A...270..494M,1993A&A...277..639S,2004A&A...422..693M}. The reduced gas pressure inside the umbra is compensated by the Lorentz force of the strong magnetic field. After a few assumptions, such as radial symmetry of the magnetic field vector, no magnetic field in the azimuthal direction and a negligible Evershed flow \citep{1977SoPh...55..335M}, this can be expressed as
\begin{align}
P_{\rm gas}(r=a,z)-P_{\rm gas}(r,z) = B_z^2(r,z)/8\pi + F_{\rm c}(r,z)/8\pi. \label{eq:press}
\end{align}
Here, $r=0$ refers to the center of the umbra and $r=a$ to a point in the quiet Sun. The second term on the right-hand side, $F_{\rm c}$ is the curvature integral:
\begin{align}
F_{\rm c} = 2 \int_r^a B_z(r',z) \frac{\partial B_r (r',z)}{\partial z} dr'. \label{eq:F_c}
\end{align}
Equation~\ref{eq:press} can be used to infer the Wilson depression of a sunspot by comparing the measured total pressure (gas pressure plus magnetic pressure) at some optical depth with the pressure stratification of a reference model atmosphere of the quiet Sun (which extends at least to the depth of the Wilson depression). The derived Wilson depression depends on the curvature integral, which cannot directly be inferred from the observations. Commonly, one assumes $F_{\rm c} = 0$ \citep[see e.g.,][]{2004A&A...422..693M}.

\begin{figure*}
\centering
\includegraphics[width=17cm]{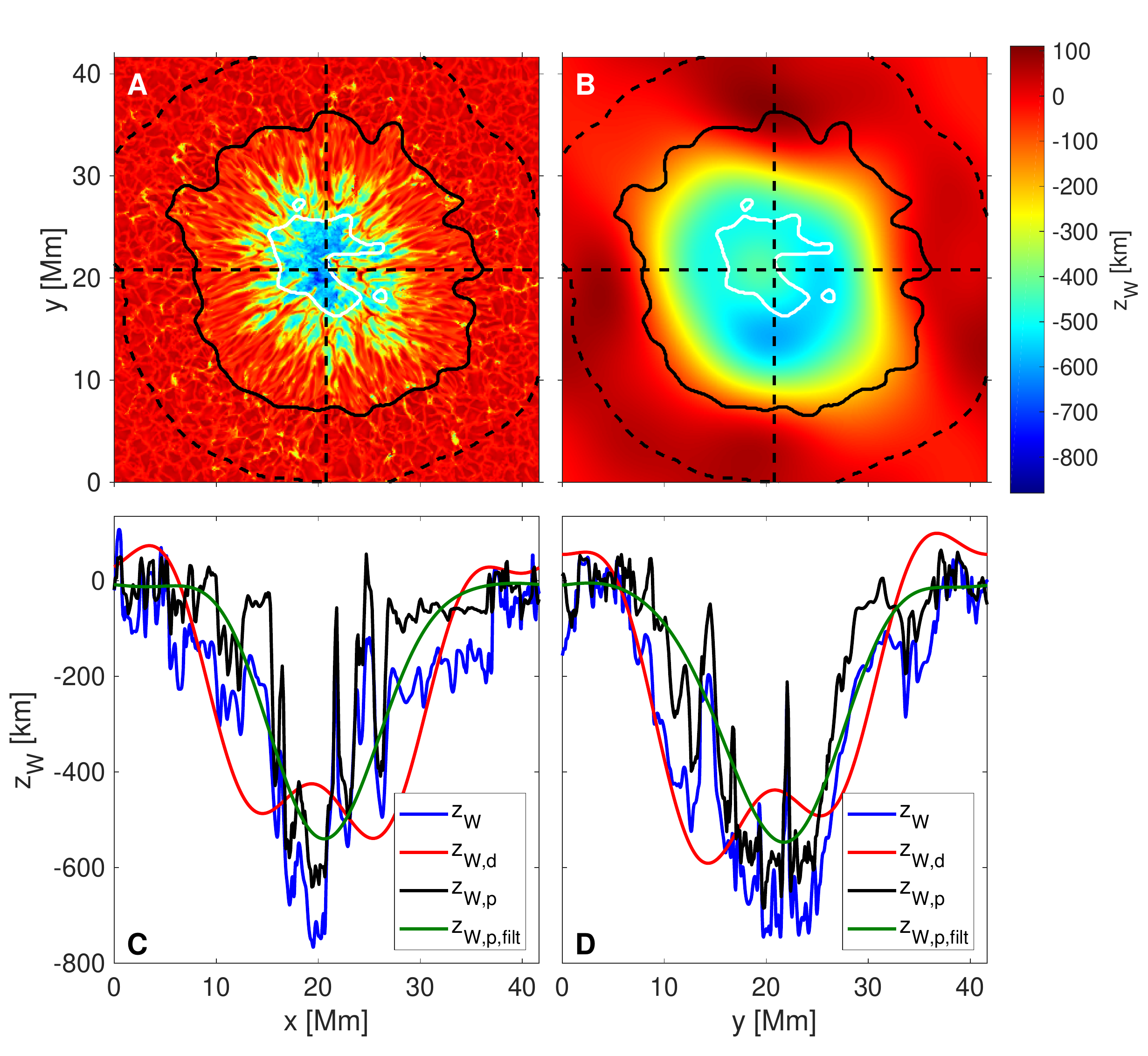}
\caption{Comparison of the Wilson depression derived from the divergence method and the pressure method for synthetic Hinode data generated from the simulation run 2. {\it Panel~A} shows the Wilson depression derived using the pressure method and {\it panel~B} the Wilson depression derived using the divergence of the magnetic field (the same image as shown in panel~B in Figure~\ref{fig:WD_spot2}). The {\it black} and {\it white} contours indicate the outer and the inner penumbral boundary, respectively. The {\it dashed black contour} shows the region that is used for defining $z_{\rm W} = 0$ (distance of at least 7~Mm to the sunspot). The {\it bottom panels} show the Wilson depression along cuts across the sunspot (marked by the {\it dashed lines} in the {\it top panels}). {\it Panel~C} shows a cut along the $x$-direction, {\it panel~D} shows a cut along the $y$-direction. {\it Blue:} true Wilson depression ($z_{\rm W}$), {\it red:} Wilson depression derived using the divergence method ($z_{\rm W,d}$), {\it black:} Wilson depression inferred from the pressure method ($z_{\rm W,p}$), {\it green:} Wilson depression inferred from the pressure method degraded to the spatial resolution of the divergence method ($z_{\rm W,p,filt}$).}
\label{fig:compare_WD}
\end{figure*}

We apply this method to both simulated sunspots. We use a horizontal average of a quiet-sun region from simulation run 2 as the reference atmosphere and extract the pressure and the magnetic field from the inversions at $\tau = 1$. The resulting Wilson depressions are in the range $420\-- 490$~km, when averaging over the umbra (see Table~\ref{tab:z_W}), in good agreement with the results from previous studies for $F_{\rm c} = 0$, which lay in the range 400\--450~km \citep{1993A&A...270..494M,1993A&A...277..639S,2004A&A...422..693M}. Figure~\ref{fig:compare_WD} compares the Wilson depression inferred from this method for simulation run 2 with our results based on the divergence of the magnetic field and the true solution. 

The pressure method results in a Wilson depression which is about 110~km smaller than the true value for the simulations. This suggests that the curvature integral is positive in both spots. We obtain the curvature integral by assuming that the difference between the true Wilson depression and the one obtained from the pressure method is due to the curvature integral neglected by the pressure method. For the spot in simulation 2, an error in the Wilson depression of 110~km corresponds to $F_{\rm c}/8\pi \approx 2.8\times 10^5 \textmd{ dyn cm}^{-2}$ in the umbra. This is about half of the value of the magnetic pressure in the umbra ($4.9 \times 10^5 \textmd{ dyn cm}^{-2}$). 

The inferred value of $F_{\rm c}/8\pi$ is in good agreement with the curvature integral derived directly from the MHD cube. The curvature integral depends on height, between $z=-700$~km and $z=-400$~km, $F_{\rm c}/8\pi$ varies between $1.5\times 10^5 \textmd{ dyn cm}^{-2}$ and $5.9\times 10^5 \textmd{ dyn cm}^{-2}$. At greater heights, it becomes negative, reaching $-1.2\times 10^6 \textmd{ dyn cm}^{-2}$ at $z=-240$~km. This change of sign is caused by the vertical derivative of the radial component of the magnetic field in Eq.~\ref{eq:F_c}. The strength of $B_r$ reaches its maximum at around $\tau = 1$ and decreases towards larger or smaller heights, causing a change of sign of its vertical derivative.

In the penumbra, the pressure method significantly underestimates the Wilson depression (by more than 100~km for the spot in simulation run 2). This is caused by using only the vertical component of the magnetic field when estimating the magnetic pressure. In the penumbra, the magnetic field is predominantly horizontal, meaning that the magnetic pressure and, hence, also the Wilson depression is underestimated when using $B_z$ only.

\subsection{Real Hinode data: case study of AR~10923} \label{sect:AR10923}
\begin{figure*}
\includegraphics[width=17cm]{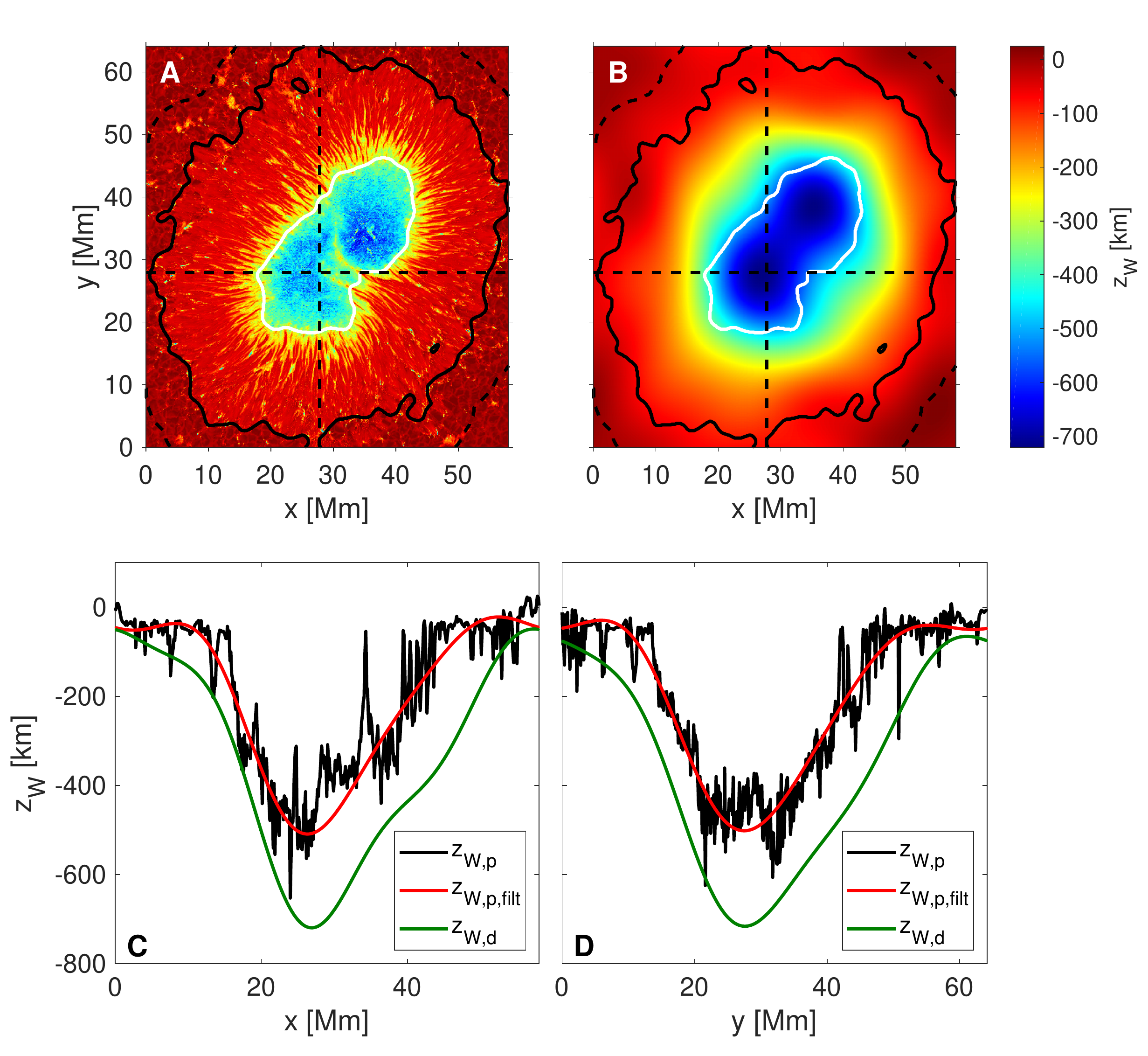}
\caption{Comparison of the Wilson depression derived from the divergence method and the pressure method for AR~10923. {\it Panel~A} shows the Wilson depression derived using the pressure method and {\it panel~B} the Wilson depression derived using the divergence method. The {\it black} and {\it white} contours indicate the outer and the inner penumbral boundary, respectively. The dashed black contour shows the region that is used for defining $z_{\rm W} = 0$ (distance of at least 7~Mm to the sunspot). The {\it bottom panels} show the Wilson depression along cuts across the sunspot (marked by the {\it dashed lines} in the {\it top panels}). {\it Panel~C} shows a cut along the $x$-direction, {\it panel~D} shows a cut along the $y$-direction. {\it Black:} Wilson depression inferred from the pressure method, {\it red:} Wilson depression inferred from the pressure method degraded to the spatial resolution of the divergence method, {\it green:} Wilson depression derived using the divergence method.}
\label{fig:WD_real_spot}
\end{figure*}

We also tested the performance of our method on the Hinode observations of AR~10923 (see Figure~\ref{fig:WD_real_spot} and Table~\ref{tab:z_W}). Again, we used a maximum dimensionless wavenumber $k_{\rm max} = l_{\rm max} = 3$, which corresponds to an effective spatial resolution of $\sim 19$~Mm in the $x$-direction and $\sim 21$~Mm in the $y$-direction. The resulting Wilson depression looks similar to the one of the simulated spots, although the average Wilson depression is by $\sim 40$~km deeper. It displays two distinct local maxima, where the Wilson depression exceeds 700~km. These maxima occur in regions with strong magnetic field (see top row of Figure~\ref{fig:inv_real_spot}). These two regions are separated by a filament, along which the Wilson depression is reduced by $\sim 60$~km. The Wilson depression smoothly decreases from the center of the umbra towards the outer penumbral boundary. The height of the $\tau = 1$ surface is more or less constant outside the sunspot, which is to be expected at the large spatial scales considered here.

Unfortunately, we cannot derive a reliable estimate of the error of the Wilson depression for this spot. As found from the simulated spots (Section~\ref{sect:simulated_spots}) the main errors in $z_{\rm W}$ arise from uncertainties in the inversions. Obtaining the errors in inverted values is never straightforward, also from the spatially coupled inversion scheme of \citet{2012A&A...548A...5V,2013A&A...557A..24V}, which was used to invert this spot. We expect that the uncertainty is similar to or larger than for the synthetic data, i. e. roughly 95~km or larger.

The Wilson depression derived using the pressure method (see Figure~\ref{fig:WD_real_spot} and Table~\ref{tab:z_W}) is significantly smaller (by about 180~km) than the one based on our method, suggesting that the curvature integral is positive in this spot, as well. Assuming the $z_{\rm W}$ deduced via the divergence method to be correct, we retrieve $F_{\rm c}/8\pi \approx 4.21 \times 10^5 \textmd{ dyn cm}^{-2}$. This is higher than the pressure from the vertical component of the magnetic field ($3.5\times 10^5 \textmd{ dyn cm}^{-2}$) in the umbra at $\tau = 1$. The high value of $F_{\rm c}$ of AR~10923 is consistent with typical values found in other spots, while the results for the simulated spots are lower than what has been reported before. \cite{1993A&A...277..639S} estimate the curvature integral to be within the range $3.5\times 10^5 \textmd{ dyn cm}^{-2} \lesssim F_{\rm c}/8\pi \lesssim 1.6\times 10^6 \textmd{ dyn cm}^{-2}$ for the sunspot that they studied. The value we find lies within this range.

\section{Discussion}
We could successfully reproduce the Wilson depression of two simulated sunspots by minimizing the divergence of the magnetic field. We also applied this method to a sunspot observed with Hinode (AR~10923) and derived a Wilson depression that is consistent with the results statistically provided by some studies of the Wilson effect \citep[$\sim 600$~km,][]{1972SoPh...26...52G}. Our results suggest that the pressure due to the vertical component of the magnetic field and the curvature integral contribute by a similar amount to the horizontal force balance within sunspots, as has already been suggested by previous investigations \citep{1993A&A...270..494M,1993A&A...277..639S,2004A&A...422..693M}. The Wilson depression of the sunspot simulations by \citet{2012ApJ...750...62R,2015ApJ...814..125R} is smaller by $\sim 50$~km  than the one of AR~10923. In the simulated spots, the curvature integral is significantly lower than the pressure due to the vertical component of the magnetic field, leading to a lower Wilson depression. This might be caused by the different geometry of the simulated spots compared to AR~10923 (see Table~\ref{tab:param}). The simulated spots exhibit a slightly stronger magnetic field but they are significantly smaller than the AR~10923 spot.

The main limitation of our method is that it requires a reliable inversion of the full magnetic field vector over a broad range in height. In the umbra, our method requires an estimate of the magnetic field up to $\sim 800$~km above the $\tau = 1$ surface. The lines used in this study are not very sensitive at these height and so, the magnetic field is predominantly extrapolated from lower layers. Observations of multiple spectral lines with a broad range of formation heights are needed in order to retrieve a better inversion (as is planned, e. g., with the upcoming Sunrise III mission). Systematic errors also arise from the assumption of LTE and hydrostatic equilibrium in the inversion, although these are likely small compared with the uncertainty resulting from the inversions. Also, modeling the height dependency of atmospheric parameters with splines is not always a good representation of the true atmosphere, especially when there are strong gradients with height. In addition, the $180^\circ$-ambiguity of the Zeemann-effect needs to be resolved accurately across the entire spot at all optical depths. Fortunately, some of these inaccuracies of the inversion occur on small spatial scales, so that their influence on large spatial scales should be limited. For example, as shown in Section~\ref{sect:method}, neglecting the small-scale corrugations of the $\tau=1$ layer has only a small influence on the large-scale divergence of the magnetic field. A detailed determination of the errors in the spatially coupled inversion is beyond the scope of this paper. We therefore use the tests made with the synthetic data as a rough guide to the total error.

The Wilson depression derived from the synthetic Hinode data exhibits an error of the order of $\sim 95$~km. In case of real observations, the error is probably somewhat higher. The synthetic data were generated using the same radiative transfer code and the same line parameters as were used for the inversion. Hence, the inversion of the synthetic data is likely to be more accurate than for real observations. Bearing these arguments in mind, we assign a preliminary error of 100~km to the Wilson depression derived from the Hinode observations.

The estimated error of our method is mainly statistical in nature. Any systematic component is significantly lower than the one of the Wilson effect or of the pressure method. As explained in the introduction, the values of the Wilson depression derived using the Wilson effect vary by more than 1000~km between different studies in a systematic manner. In case of the pressure method, the inferred Wilson depression depends on the assumed value of the curvature integral.

Our method is limited to the large-scale corrugations of the $\tau = 1$ surface. Measuring the Wilson depression with a higher spatial resolution requires including more Fourier coefficients in the Fourier series of the Wilson depression, which affects the minimization of the merit function. In addition, the divergence of the magnetic field is dominated by the largest spatial scales. When evaluating the divergence of the magnetic field for AR~10923 on a corrugated grid in geometrical height (300~km above $\tau = 1$ at each pixel), 76\% of the total variance of the divergence of the magnetic field occur within the range of spatial scales that we considered in Section~\ref{sect:AR10923} ($k_{\rm max} = l_{\rm max} = 3$). Hence, the merit function is not very sensitive to smaller spatial scales, at least when considering entire sunspots.

\begin{acknowledgements}
This work benefited from the Hinode sunspot database at MPS, created by Gautam Narayan. This project has received funding from the European Research Council (ERC) under the European Union’s Horizon 2020 research and innovation programme (grant agreement No 695075) and has been supported by the BK21 plus program through the National Research Foundation (NRF) funded by the Ministry of Education of Korea.
\end{acknowledgements}

\bibliographystyle{aa} 
\bibliography{literature} 

\begin{thebibliography}{22}
\expandafter\ifx\csname natexlab\endcsname\relax\def\natexlab#1{#1}\fi

\bibitem[{{Balthasar} \& {W{\"o}hl}(1983)}]{1983SoPh...88...71B}
{Balthasar}, H. \& {W{\"o}hl}, H. 1983, \solphys, 88, 71

\bibitem[{{Frutiger} {et~al.}(2000){Frutiger}, {Solanki}, {Fligge}, \&
  {Bruls}}]{2000A&A...358.1109F}
{Frutiger}, C., {Solanki}, S.~K., {Fligge}, M., \& {Bruls}, J.~H.~M.~J. 2000,
  \aap, 358, 1109

\bibitem[{{Georgoulis}(2005)}]{2005ApJ...629L..69G}
{Georgoulis}, M.~K. 2005, \apjl, 629, L69

\bibitem[{{Gokhale} \& {Zwaan}(1972)}]{1972SoPh...26...52G}
{Gokhale}, M.~H. \& {Zwaan}, C. 1972, \solphys, 26, 52

\bibitem[{{Ichimoto} {et~al.}(2008){Ichimoto}, {Lites}, {Elmore}, {Suematsu},
  {Tsuneta}, {Katsukawa}, {Shimizu}, {Shine}, {Tarbell}, {Title}, {Kiyohara},
  {Shinoda}, {Card}, {Lecinski}, {Streander}, {Nakagiri}, {Miyashita},
  {Noguchi}, {Hoffmann}, \& {Cruz}}]{2008SoPh..249..233I}
{Ichimoto}, K., {Lites}, B., {Elmore}, D., {et~al.} 2008, \solphys, 249, 233

\bibitem[{{Kosugi} {et~al.}(2007){Kosugi}, {Matsuzaki}, {Sakao}, {Shimizu},
  {Sone}, {Tachikawa}, {Hashimoto}, {Minesugi}, {Ohnishi}, {Yamada}, {Tsuneta},
  {Hara}, {Ichimoto}, {Suematsu}, {Shimojo}, {Watanabe}, {Shimada}, {Davis},
  {Hill}, {Owens}, {Title}, {Culhane}, {Harra}, {Doschek}, \&
  {Golub}}]{2007SoPh..243....3K}
{Kosugi}, T., {Matsuzaki}, K., {Sakao}, T., {et~al.} 2007, \solphys, 243, 3

\bibitem[{{Lites}(2007)}]{2007ASPC..369..579L}
{Lites}, B.~W. 2007, in Astronomical Society of the Pacific Conference Series,
  Vol. 369, New Solar Physics with Solar-B Mission, ed. K.~{Shibata},
  S.~{Nagata}, \& T.~{Sakurai}, 579

\bibitem[{{Lites} {et~al.}(2013){Lites}, {Akin}, {Card}, {Cruz}, {Duncan},
  {Edwards}, {Elmore}, {Hoffmann}, {Katsukawa}, {Katz}, {Kubo}, {Ichimoto},
  {Shimizu}, {Shine}, {Streander}, {Suematsu}, {Tarbell}, {Title}, \&
  {Tsuneta}}]{2013SoPh..283..579L}
{Lites}, B.~W., {Akin}, D.~L., {Card}, G., {et~al.} 2013, \solphys, 283, 579

\bibitem[{{Maltby}(1977)}]{1977SoPh...55..335M}
{Maltby}, P. 1977, \solphys, 55, 335

\bibitem[{{Mart{\'i}nez Pillet} \& {V{\'a}zquez}(1993)}]{1993A&A...270..494M}
{Mart{\'i}nez Pillet}, V. \& {V{\'a}zquez}, M. 1993, \aap, 270, 494

\bibitem[{{Mathew} {et~al.}(2004){Mathew}, {Solanki}, {Lagg}, {Collados},
  {Borrero}, \& {Berdyugina}}]{2004A&A...422..693M}
{Mathew}, S.~K., {Solanki}, S.~K., {Lagg}, A., {et~al.} 2004, \aap, 422, 693

\bibitem[{{Prokakis}(1974)}]{1974SoPh...35..105P}
{Prokakis}, T. 1974, \solphys, 35, 105

\bibitem[{{Puschmann} {et~al.}(2010){Puschmann}, {Ruiz Cobo}, \&
  {Mart{\'{\i}}nez Pillet}}]{2010ApJ...720.1417P}
{Puschmann}, K.~G., {Ruiz Cobo}, B., \& {Mart{\'{\i}}nez Pillet}, V. 2010,
  \apj, 720, 1417

\bibitem[{{Rempel}(2012)}]{2012ApJ...750...62R}
{Rempel}, M. 2012, \apj, 750, 62

\bibitem[{{Rempel}(2015)}]{2015ApJ...814..125R}
{Rempel}, M. 2015, \apj, 814, 125

\bibitem[{{Solanki}(1987)}]{1987PhDT.......251S}
{Solanki}, S.~K. 1987, {Ph.D. Thesis No. 8309} (ETH, Zürich)

\bibitem[{{Solanki}(2003)}]{2003A&ARv..11..153S}
{Solanki}, S.~K. 2003, \aapr, 11, 153

\bibitem[{{Solanki} {et~al.}(1993){Solanki}, {Walther}, \&
  {Livingston}}]{1993A&A...277..639S}
{Solanki}, S.~K., {Walther}, U., \& {Livingston}, W. 1993, \aap, 277, 639

\bibitem[{{van Noort}(2012)}]{2012A&A...548A...5V}
{van Noort}, M. 2012, \aap, 548, A5

\bibitem[{{van Noort} {et~al.}(2013){van Noort}, {Lagg}, {Tiwari}, \&
  {Solanki}}]{2013A&A...557A..24V}
{van Noort}, M., {Lagg}, A., {Tiwari}, S.~K., \& {Solanki}, S.~K. 2013, \aap,
  557, A24

\bibitem[{{V{\"o}gler} {et~al.}(2005){V{\"o}gler}, {Shelyag}, {Sch{\"u}ssler},
  {Cattaneo}, {Emonet}, \& {Linde}}]{2005A&A...429..335V}
{V{\"o}gler}, A., {Shelyag}, S., {Sch{\"u}ssler}, M., {et~al.} 2005, \aap, 429,
  335

\bibitem[{{Wilson} \& {Maskelyne}(1774)}]{1774RSPT...64....1W}
{Wilson}, A. \& {Maskelyne}, N. 1774, Philosophical Transactions of the Royal
  Society of London Series I, 64, 1

\end{thebibliography}

\end{document}